\documentclass[showpacs,aps]{revtex4}
\usepackage{graphicx}
\usepackage{amsmath,amssymb,subfigure}
\usepackage{color}

\def\bc{\begin{center}}
\def\ec{\end{center}}

\def\beq{\begin{equation}}
\def\eeq{\end{equation}}

\begin{document}

\title{Inelastic processes in K$^{+}-$He collisions in energy range $0.7-10$
keV}
\author{R.A. Lomsadze$^{1}$, M.R. Gochitashvili$^{1}$, R.Ya. Kezerashvili$%
^{2,3}$, N.O. Mosulishvili$^{1}$, and R. Phaneuf$^{4}$}
\affiliation{$^{1}$Tbilisi State University,
Tbilisi, 0128, Republic of Georgia\\
$^{2}$New York City College of Technology, The City
University of New York, Brooklyn, NY 11201, USA \\
$^{3}$The Graduate School and University Center, The City University of New
York, New York, NY 10016, USA \\
$^{4}$Department of Physics, University of Nevada, Reno, NV 89557, USA }
\date{\today}

\begin{abstract}
Absolute cross sections for charge exchange, ionization, stripping and
excitation in K$^{+}-$He collisions were measured in the ion energy range $%
0.7-10$ keV. The experimental data and the schematic correlation diagrams
are used to analyze and determine the mechanisms for these processes. The
increase of the excitation probability of inelastic channels with the angle
of scattering is revealed. An exceptionally highly excited state of He is
observed and a peculiarity for the excitation function of the resonance line
is explained. The intensity ratio for the excitation of the K II $\lambda
=60.1$ nm and $\lambda =61.2$ nm lines is 5:1 which indicates the high
probability for excitation of the singlet resonance level $^{1}$P$_{1}$
compared to the triplet level $^{3}$P$_{1}$. The similarity of the
population of the 4p state of the potassium ion and atom as well as the
anomalously small values of the excitation cross sections are explained.
\end{abstract}

\maketitle

\section{Introduction}

Ion-atom collisions have been an attractive subject and are of considerable
interest in atomic physics due to both their importance in fundamental
physics and their application in many fields, such as laboratory and
astrophysical plasmas \cite{Nikit}, heavy ion inertial fusion \cite{Logan},
radiation physics, collisional and radioactive processes in the Earth's
upper atmosphere \cite{Betz, Kez} and many other technological areas. In
recent decades, ion-atom collisions have been studied in detail
experimentally as well as theoretically from low to relativistic collision
energies (see, for example, Refs. \cite{Nikit, Betz, Moshammer, Kog,
Kaganovich}). There has been an increased need the evaluation of ion-atom
cross sections of different processes for many accelerator applications \cite%
{Shevelko}. For example, a beam interaction with the remaining background
gas and gas desorbed from walls limits the intensity of bunches at the RHIC
(Relativistic Heavy Ion Collider) \cite{Fischer} and a pressure rise from
ion losses at the low-energy antiproton ring brought concerns for the LHC
(Large Hadron Collider) \cite{Demma}.

At moderate energies, collisions of closed-shell ions with closed-shell
atoms for various inelastic channels such as the ionization,
charge-exchange, stripping, and excitation are well understood. However, the
absolute cross section information of these inelastic processes are of
interest - in particular, the energy dependence of collisions of K$^{+}$
ions with He atoms because the excitation probabilities in these collisions
(with some exceptions) are one or two orders of magnitude smaller than in
symmetric and quasisymmetric systems \cite{Kikiani, Afrosimov1, Afrosimov2,
Hegerberg}. Another remarkable feature of inelastic collisions of these
closed-shell particles is that the excitation into a doubly excited state
occurs with a high probability. Therefore, an accurate determination of the
structure of different inelastic cross sections for these collisions is
important to understand the mechanisms for inelastic transitions in the
outer and, in some cases, inner shells of colliding atomic particles. In
order to discuss quantitatively the excitation mechanism one has to evaluate
the parameters of existing theories and explore the contributions of
separate inelastic channels for the investigated processes by using
experimental measurements.

Despite many studies of K$^{+}-$He collisions which have been carried out by
various methods \cite{Kikiani, Afrosimov2, Hegerberg, Ogurtsov, Moe, Bidin,
Matveev, Matveev2, Mouzon, Kita}, available data for the absolute cross
sections of the above mentioned processes are contradictory \cite{Kikiani,
Moe, Bidin, Mouzon} and, in some cases, unreliable \cite{Ogurtsov}.

Charge exchange cross sections for K$^{+}-$Ar, K$^{+}-$Kr, K$^{+}-$Xe and K$%
^{+}-$He systems were reported in Ref. \cite{Ogurtsov} using the detection
of fast neutral particles within a definite interval of scattering angles.
However, as shown in Ref. \cite{Afrosimov3} the restriction on the interval
for collision angles in Ref. \cite{Ogurtsov} underestimated the measured
charge exchange cross sections by a factor of ten for the K$^{+}-$Ar, Kr, Xe
collision pairs over the entire energy range considered. The charge exchange
cross sections reported in Ref. \cite{Ogurtsov} might also have been
underestimated for K$^{+}$ - He collisions. Hence, it was necessary to carry
out measurements of the charge-exchange cross sections for K$^{+}-$He
collisions in a wider and more complete interval of scattering angles using
a method that is free of this deficiency.

In Ref. \cite{Kikiani4}, only the absolute value of the cross section for
excitation processes, occurring in the case when the mass of the incident
ion is less than of target particle ( Na$^{+}-$Ar) was reported. The results
of the measurements of the excitation function in arbitrary units for Na$%
^{+}-$He and K$^{+}-$He collision pairs are presented in Refs. \cite%
{Matveev, Matveev2}. The lack of absolute excitation cross sections for the
asymmetric K$^{+}-$He system motivated the present detailed investigation of
the primary mechanisms for this collision process.

In order to make use of the measured cross section for electron production,
it is necessary to know absolute ionization cross sections for the target
atoms and absolute stripping cross sections for the incident ions. Since
there are no previous measurements of the stripping cross section for K$%
^{+}- $He collisions, it is imperative to measure this cross section.

As is known, the basic mechanism responsible for the process of ionization
in atomic collisions is a diabatic orbital super promotion into the
continuum. Proposed in Ref. \cite{Dem}, this mechanism was further developed
in Refs. \cite{Solov, Solov2, PhysRep}. Advantages and limitations of these
studies are assessed for colliding systems with one \textquotedblleft
active\textquotedblright\ electron, e.g. H$^{+}-$H and He$^{2+}-$H. It is
important to extend the theory to ionization processes in collisions of
many--electron atomic systems. Experimental methods \cite{Woer, Zinov, Ogur2}
that are used so far to study this mechanism are based on the measurement of
continuous parts of the energy spectra of ejected electrons. More detailed
information for the ejected electrons at different impact parameters can be
obtained using coincidence techniques.

One of the objectives of this work is to show that in many cases the
information extracted from complicated coincidence experiments can also be
obtained using a simple method, namely by measuring the energy loss spectra
of incident particles. Such spectra can be measured over a wide range of
scattering angles and energy losses, for which the problems of collection of
low energy electrons do not arise. Below, we report absolute total cross
sections for charge exchange processes, ionization, and stripping processes
that result in the production of free electrons, as well as the excitation
of both the projectile and target particles and energy loss spectra in
collisions of K$^{+}$ ions with He atoms.

The remainder of this paper is organized in the following way. In Sec.~\ref%
{Exper} we introduce our unique experimental setup that includes three
collision chambers and present the experimental techniques and procedures
used to measure the absolute total cross sections for the excitation, charge
exchange, ionization, and stripping processes. Results of measurements and
discussion of mechanisms for different processes occurring in K$^{+}-$He
collisions are given in Sec. \ref{results}. Finally, in Sec.~\ref{Concl} we
summarize our studies and present the conclusions.

\section{Experimental technique}

\label{Exper}

A schematic view of the experimental setup is shown in Fig. 1. A beam of K$%
^{+}$ ions from a surface -- ionization ion source is accelerated and
focused by an ion-optics system, which includes quadruple lenses and
collimating slits. After the beam passes through a magnetic mass
spectrometer, it enters the first collision chamber containing He gas for
measurements of excitation processes in K$^{+}-$He collisions. Our setup is
designed so that when the beam of K$^{+}$ ions passes through the second
collision chamber measurements of the charge exchange, ionization and
stripping processes occur. The energy loss spectrum is obtained by a
collision spectroscopy method when K$^{+}$ ions collide with target He atoms
in the third collision chamber. The pressure in each collision chamber when
there is no He target gas is kept below 10$^{-6}$ Torr and the typical
pressure under operation is 10$^{-4}$ Torr. The measurements are performed
under a single-collision conditions. The current of primary ions in the
collision chamber is $I$=0.01-0.1 $\mu $A. The uniqueness of our
experimental approach is that the quality of the beam as well as the
experimental conditions for all processes under investigation always remain
identical. Moreover, the experimental techniques include a condenser plate
method, angle- and energy-dependent collection of product ions, energy-loss
and optical spectroscopy under the same "umbrella". The measurement
procedures have been discussed in some detail previously \cite{Kikiani3,
Kikiani4, Gochit, Gocht2}, therefore the description of the present
measurements can be kept comparatively short.

\subsection{The first collision chamber}

\begin{figure}[th]
\centering {\ \includegraphics[width=0.48\textwidth]{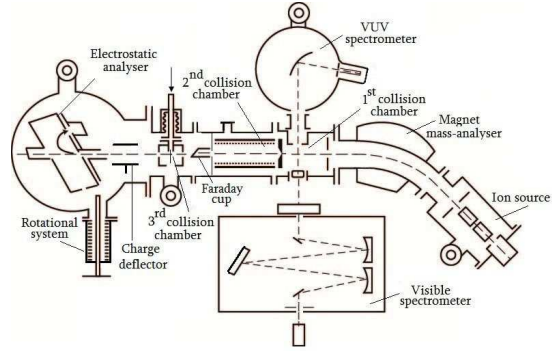} }
\caption{Schematic diagram of the experimental set up.}
\label{Fig1}
\end{figure}
The measurements of the excitation processes in K$^{+}-$He collision were
performed when the beam of K$^{+}$ ions passed through the first collision
chamber. Cross sections for the excitation processes were measured by the
optical spectroscopy method \cite{Gocht2}. The radiation emitted in the
first collision chamber was observed at an angle of 90$^{0}$ with respect to
the direction of the primary ion beam. The spectral analysis of this
radiation was performed in the vacuum ultraviolet (VUV) as well as visible
spectral regions. An electron multiplier was used to detect the intensity of
the radiation. Particular attention was devoted to the reliable
determination and control of the relative and absolute spectral sensitivity
of the light-recording system. This was done by measuring the signal due to
the emission of molecular bands and atomic lines excited by electrons in
collisions with H$_{2}$, N$_{2}$, O$_{2}$ and Ar. For this, an electron gun
was placed directly in front of the entrance slit of the collision chamber.
The relative spectral sensitivity, and the values of the absolute cross
sections, were obtained by comparing the cross sections for the same lines
and molecular bands reported in the literature \cite{Ajello, Mumma, Stone,
Tan, Tan2}. The uncertainties in the excitation cross sections for the K $%
^{+}-$He system are estimated to be 20\% and the uncertainty of the relative
measurements does not exceed 5\%.


\subsection{The second collision chamber}

When the K$^{+}$ beam enters the second collision chamber filled with the
target gas, the measurements for the charge exchange and ionization
processes occur. The charge exchange and ionization cross sections were
measured by a refined version of the capacitor method \cite{Kikiani3}. In an
earlier paper \cite{Ogurtsov} the measurements were performed by the
standard transfer electric field method. The customary procedure is to use
one of the central electrodes as the measurement electrode. We consider that
such approach is the reason for significant errors in measurements \cite%
{Ogurtsov} because scattered K$^{+}$ ions may strike the electrodes used for
measurements. This effect becomes more evident especially for low energy
collisions. To avoid this deficiency we accordingly used a refined version
of the transfer electric field method in which the effect of scattering on
the measured results is substantially reduced by shifting from the central
electrode (standard method) to the first electrode (towards beam entrance
side). Due to fringing effects at the edges of this electrode a system of
auxiliary electrodes between the first electrode and the entrance slit was
installed. These auxiliary electrodes create a uniform potential near the
first electrode. The first electrode, the auxiliary electrodes, and the
entrance slit are all positioned as close together as possible. This close
arrangement limits the scattering region on the beam -- entrance side. Thus,
according to our estimation, taking into account the geometry of our
facilities, only those ions which are scattered through angles greater than
70$^{0}$ can reach the electrodes.


The primary ions are detected by the Faraday cup. Collision particles
(secondary positive ions and free electrons) are detected by a collector.
The collector consists of two rows of plate electrodes that run parallel to
the primary ion beam. A uniform transverse electric field, responsible for
the extraction and collection of collision particles, is created by the
potentials applied to the grids. This method yields direct measurements of
the cross sections $\sigma ^{+}$ for the production of singly positively
charged ions and $\sigma ^{-}$ for electrons as the primary beam passes
through the gas under study. These measured quantities are related in an
obvious way to the capture cross section $\sigma _{c}$ and the apparent
ionization cross section $\sigma _{i}$, and are determined as

\begin{equation}
\sigma ^{+}=\sigma _{c}+\sigma _{i},\ \ \ \ \ \ \ \ \ \ \sigma ^{-}=\sigma
_{s}+\sigma _{i}.  \label{CrSec}
\end{equation}%
In \eqref{CrSec} $\sigma _{s}$ is the cross section stripping of the
incident ion. The cross section $\sigma _{i}$ is always larger than the
cross section for stripping $\sigma _{s}.$ It should be mentioned that if
the cross sections for multiple ionization and ionization with capture are
small then $\sigma _{i}\approx \sigma ^{-}$.

The uncertainty in the measurements of the absolute values of the cross
sections $\sigma ^{+}$ and $\sigma ^{-}$ is estimated to be 15\% over the
entire energy interval studied. This is determined primarily by the
uncertainty in the measurement of target gas pressure in the collision
chamber. The uncertainty in the measurement of the ionization cross section $%
\sigma _{i}$ is estimated to be $\sim 20\%$ over the entire energy range.
The larger value in comparison with that for the measurement of the cross
section $\sigma ^{-}$ is a consequence of the presence of an additional
error in the measurements of $\sigma _{i}$ due to stripping of K$^{+}$ ions.
At K$^{+}$ energies less than 2.0 keV the cross section $\sigma ^{+}$ is
significantly larger than $\sigma ^{-}$. Accordingly, the error in the
determination of the capture cross section $\sigma _{c}$ in this energy
region is determined primarily by the error in the measurement of $\sigma
^{+}$. With increasing K$^{+}$ energy the cross sections $\sigma ^{+}$ and $%
\sigma ^{-}$ become more nearly equal. As a result the error in the
determination of $\sigma _{c}$ increases. At a K$^{+}$ energy of 5 keV it is
estimated to be 25\%.

The stripping cross sections were measured in an independent experiment. A
beam of K$^{+}$ ions from a surface-ionization ion source after acceleration
and focusing passes through the second collision chamber. It is then
analyzed for the charge composition by an additional 90$^{0}$ magnetic mass
spectrometer (not shown in Fig.1). A set of slits that were located in front
of the detector enable the profile of the projectile beam to be to scanned
in detail. The ratio of the number of doubly charged ions, produced as a
result of stripping, to the incident ion beam, K$^{2+}/$K$^{+},$ is
determined from the areas under the corresponding lines in the mass
spectrum. Since inelastic processes, apparently including stripping, may be
accompanied by the scattering of particles through comparatively large
angles in collisions between alkali metal ions and inert gas atoms \cite%
{Afrosimov1, Afrosimov2, Afrosimov3}, special measures were taken to arrange
complete collection of K$^{++}$ ions.

\subsection{The third collision chamber}

The energy loss spectrum was obtained by the collision spectroscopy method
\cite{Gochit}. The primary beam extracted from the ion source was
accelerated to the desired energy before being analyzed according to $q/m$ ($%
q$ and $m$ are the ion's charge and mass, respectively). The analyzed ion
beam was then allowed to pass through the third collision chamber by
appropriately adjusting the slits prior to entering into a `box' type
electrostatic analyzer. The energy resolution of this analyzer is $\Delta
E/E $=1/500. Automatic adjustments of analyzer potentials gave the
possibility for investigation of energy loss spectra in the energy range of
0 -- 100 eV. The differential cross section is measured by rotating the
analyzer around the center of collisions over an angular range between 0 and
25$^{0}$ degrees. The laboratory angle is determined with respect to the
primary ion beam axis with an accuracy of 0.2$^{0}.$

For the measurements of the charge exchange differential cross section the
charge component of scattered primary particles realized in the third
collision chamber is separated by the electric field (see ion charge
deflector in Fig. 1) and neutral particles formed by electron capture
collisions are registered by the secondary electron multiplier. Such a tool
gives us the possibility to determine total cross sections and compare them
with the results obtained in the second collision chamber. In addition, the
measured energy loss spectrum gives detailed information related to the
intensity of inelastic processes realized in the first collision chamber
(excitation) and in the second collision chamber (charge exchange,
ionization and stripping).

\section{Results of Measurements and Discussion}

\label{results}

The collision of a K$^{+}$ ion beam with He gas leads mainly to the
following processes: charge exchange reaction

\begin{equation}
\text{K}^{\text{+}}+\text{He}\rightarrow \text{K}+\text{He}^{\text{+}},
\label{CE}
\end{equation}%
when the K atom and He$^{+}$ ion products can be in the ground state or in
different excited states; different excitation processes

\begin{equation}
\text{K}^{\text{+}}+\text{He}\rightarrow \text{K}^{+}+\text{He},
\label{K excited}
\end{equation}%
that include different channels for excitation of the K$^{+}$ ion or/and He
atom; stripping of K$^{+}$

\begin{equation}
\text{K}^{\text{+}}+\text{He}\rightarrow \text{K}^{++}+\text{He}+e
\label{striping}
\end{equation}%
and ionization of He

\begin{equation}
\text{K}^{\text{+}}+\text{He}\rightarrow \text{K}^{+}+\text{He}^{+}+e
\label{ioniz}
\end{equation}%
as well as excitation of autoionization states

\begin{equation}
\text{K}^{\text{+}}+\text{He}\rightarrow \text{K}^{+}+\text{He}^{\ast \ast
}\rightarrow \text{K}^{+}+\text{He}^{+}+e  \label{auto He}
\end{equation}%
when He$^{\ast \ast }$ is formed in different autoionization states and then
decays by emitting an electron.

In Figs. 2 -- 4 the measured energy dependences of absolute total
cross-sections are presented for charge exchange into the ground and excited
states of K$^{+}$, the ionization, stripping and excitation function of
potassium atom, ion and target helium atom in K$^{+}-$ He collisions. Figs.
5b $-$ 5d represent a typical example of the energy loss spectrum for K$%
^{+}- $He system. For comparison, in Fig. 5a is given the energy loss
spectrum for K$^{+}-$Ar collision. We have also measured the energies of the
electrons emitted in the collisions. It was found that the energy of most
liberated electrons is below 10-15 eV.

\begin{figure}[ht]
\centering {\ \includegraphics[width=0.48\textwidth]{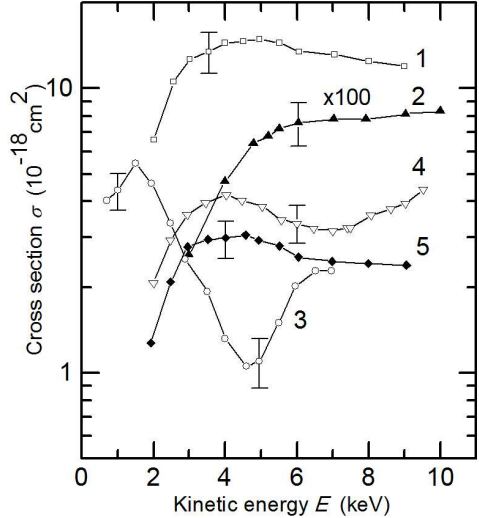} } \vspace{%
-0.5cm}
\caption{Dependencies of excitation cross sections and total charge exchange
cross sections on energy of the K$^{+}$ ion in K$^{+}-$He collisions.
Curves: (1) - K II ($\protect\lambda =60.1 $ nm) for process (3); (2) - K I (%
$\protect\lambda =766.5$ nm) for process (2); (3) - charge exchange process
when K is in the ground state in process (2); (4) - He I ($\protect\lambda %
=58.4$ nm) for process (3); (5) K II ($\protect\lambda =61.2$ nm) for
process (3).}
\label{Fig2}
\end{figure}

As seen from Figs. 2 and 3, a distinctive feature of most of these inelastic
processes is the small magnitude of the cross section. The excitation
functions of the K II lines at $\lambda =60.1$ nm and $\lambda =61.2$ nm are
similar (curve 1 and 5 in Fig. 2). The curves did not differ in shape and
curve 1 could be obtained from curve 5 by multiplying the ordinate of the
last one by a factor of $\sim $ 5. Another feature that is prominent in the
ionization, charge exchange and excitation cross sections is the difference
in the energy dependences. While the ionization cross section (Fig. 3) and
excitation function of the K$^{+}$ ion (Fig. 2, curve 5) and K atom (Fig. 2
curve 2) are small at low energies and increase monotonically with energy up
to 5 keV, the total charge exchange cross section has a complex energy
dependence and generally decreases with the collision energy up to 5 keV and
then increases (Fig. 2, curve 3). Structural features are also observed in
the resonance line for the helium atom (Fig. 2, curve 4).

\begin{figure}[ht]
\centering {\ \includegraphics[width=0.48\textwidth]{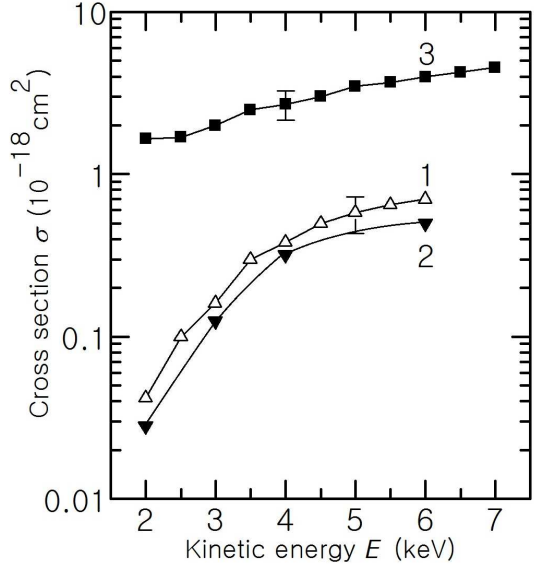} }
\caption{The absolute cross section for K$^{+}+$ He collisions. The
formation of K$^{++}$ product ions is shown by curve 1. Results of
calculations of the production of K$^{++}$ in K$^{+}+$ He collisions from
Ref. \protect\cite{Solov} are indicated by curve 2. Curve 3 indicates the
free electron production in K$^{+}+$ He collision.}
\label{Fig3}
\end{figure}

\begin{figure}[th]
\centering {\ \includegraphics[width=0.48\textwidth]{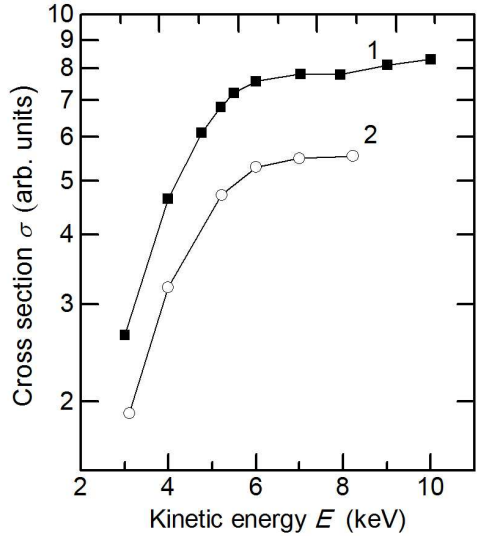} }
\caption{The excitation functions for the potassium atom K and ion K$^{+}$
in K$^{+}-$He collisions. (1) - KI, $\protect\lambda =766.5$ nm. (2) - KII, $%
\protect\lambda =389.8$ nm. }
\label{Fig4}
\end{figure}

Distinct features are observed in the energy loss spectra shown in Figs. 5.
In the case for which colliding particles have nearly equal masses (K$^{+}-$%
Ar , Fig. 5 (a) ) the main inelastic process is realized effectively in the
10 $-$ 35 eV energy loss interval, while for asymmetric colliding pairs (K$%
^{+}-$He, Fig. 5d) this energy losses are extended up to 60 eV.

The data obtained in this study can be used to draw certain conclusions
about the possible reasons for these features in the cross sections and
mechanisms for the corresponding processes. To explain these mechanisms we
use a schematic correlation diagram of the adiabatic quasimolecular terms
for the system of colliding particles. The diagram was constructed based on
Barat-Lichten rules \cite{Barat} and is presented in Fig. 6.

\subsection{Excitation functions}

These results motivate us to explore the excitation functions for collisions
of K$^{+}$ ions with helium atoms. The radiation generated in these
collisions is shown in Fig. 2 and it includes: one resonance line of the K
atom at $\lambda =766.5$ nm (curve 2) corresponding to the channel K$^{+}$(3p%
$^{6}$) $+$ He(1s$^{2}$) $\rightarrow $K(4s)$+$He$^{+}$(1s) of process (\ref%
{CE}), two resonance lines of the K$^{+}$ ion $\ $at $\lambda =60.1$ nm
(curve 1) and $\lambda =61.2$ nm (curve 5) for process (\ref{K excited}) and
one resonance line of the He atom at $\lambda =58.4$ nm (curve 4)
corresponding to different channels of process (\ref{K excited}).

The KII $\lambda =60.1$ nm line representing a transition from the $^{1}$P$%
_{1}$(3s$^{2}$3p$^{5}$ [$^{2}$P$_{1/2}$]$^{0}$4s$^{^{\prime }}$) level to
the ground state $^{1}$S$_{0}$(3s$^{2}$3p$^{6}$) is much stronger than the $%
\lambda =61.2$ nm line, corresponding to transitions from the state $^{3}$P$%
_{1}$(3s$^{2}$3p$^{5}$[$^{2}$P$_{3/2}$]$^{0}$4s) to the ground state. The
intensity ratio of the K II $\lambda =60.1$ nm and $\lambda =61.2$ lines
shown in Fig. 2 is 5:1, indicating that the probability of excitation of the
singlet resonance level $^{1}$P$_{1}$ is greater than that of the triplet
level $^{3}$P$_{1}$, due to spin conservation.

Curve 4 in Fig. 2 for the excitation function of the He(1s2p) resonance line
for the channel K$^{+}$(3p$^{6}$) $+$ He(1s$^{2}$) $\rightarrow $K$^{+}$(4p)$%
+$He(1s2p) has unusual energy dependence. This results from the overlapping
of different excitation channels in different energy ranges. Particularly,
the analyses of the correlation diagram in Fig. 6 leads to the conclusion
that at relatively small energies the excitation of He(1s2p) originate due
to the exchange interaction through the intermediate state corresponding to
the channel K$^{+}$(3p$^{6}$) $+$ He(1s$^{2}$)$\rightarrow $K(4s) $+$He$^{+}$%
(1s), while at large energies the excitation of He(1s2p) may occur due to
the $\Sigma -\Pi $ transition between terms which correspond to the K$^{+}$%
(3p$^{6}$) $+$ He(1s$^{2}$)$\rightarrow $K$^{+}$(4s$^{^{\prime }}$)$+$He(1s$%
^{2}$) and K$^{+}$(3p$^{6}$) $+$ He(1s$^{2}$)$\rightarrow $K$^{+}$(3p$^{6}$)$%
+$He(1s2p) channels.

\subsection{The charge exchange reaction}

In determining the processes responsible for charge exchange in K$^{+}-$He
collisions, we compare the total charge exchange cross sections presented in
Fig. 2 by curve 3 with those corresponding to the decay of resonant levels
of the potassium atom presented by curve 2. Taking into account the
selection rules and the ratio of oscillator strengths for the transitions,
we show that the decay of any of the excitation levels of the potassium atom
culminates in about half the cases with a transition of the atom to a
resonant state, which then decays. According to our data, the decay cross
section of the resonant levels of the potassium atom in K$^{+}$- He
collisions increases with incident ion energy, reaching about $2\times
10^{-20}$ cm$^{2}$ at an ion energy of 3 keV and 7$\times $10$^{-20}$ cm$%
^{2} $ at 5-7 keV. At low collision energies these cross sections are less
than 1\%, and at high collision energies are $\sim 5\%$ of the total charge
exchange cross sections. It follows that in these collisions the cross
section for the capture of an electron to an excited state of atom through
the channels K$^{+}$(3p$^{6}$) $+$ He(1s$^{2}$) $\rightarrow $ K(4p)$+$He$%
^{+}$(1s) or K$^{+}$(3p$^{6}$) $+$ He(1s$^{2}$)$\rightarrow $K(3d)$+$He$^{+}$%
(1s) is less than 10\% of the total charge exchange cross section. In other
words, the charge exchange process results primarily from capture of an
electron to the ground state of the K atom. The target ions are produced in
these processes primarily in the ground state. Processes involving capture
accompanied by excitation of the resultant target ion He$^{+}$ make a small
contribution into the charge exchange cross section, apparently because the
energy defect of these processes is of the same order of magnitude as the
collision energy (the total kinetic energy of the particles in the c.m.
system) \cite{McD}. The energy defect of these processes is $\sim $61 eV,
while the collision energy at an ion energy of 700 eV is only 65 eV in the
c.m. system. We thus can conclude that the charge exchange reaction (\ref{CE}%
) is governed primarily by the capture of an electron in the ground state of
the potassium atom, accompanied by the formation of a helium ion also in the
ground state: K$^{+}$(3p$^{6}$) $+$ He(1s$^{2}$) $\rightarrow $K(4s) $+$ He$%
^{+}$(1s). This process can occur, as can be seen from the diagram in Fig.
6, as a result of the direct pseudocrossing of the term corresponding to the
state K(4s) $+$ He$^{+}$(1s) with the ground state of the system. Since the
K (4s) $+$ He$^{+}$(1s) state has only $\Sigma $ symmetry, it follows that $%
\Sigma -\Sigma $ transitions play a dominant role in the charge exchange
processes.

\begin{figure}[t]
\begin{center}
\includegraphics[height=6.cm]{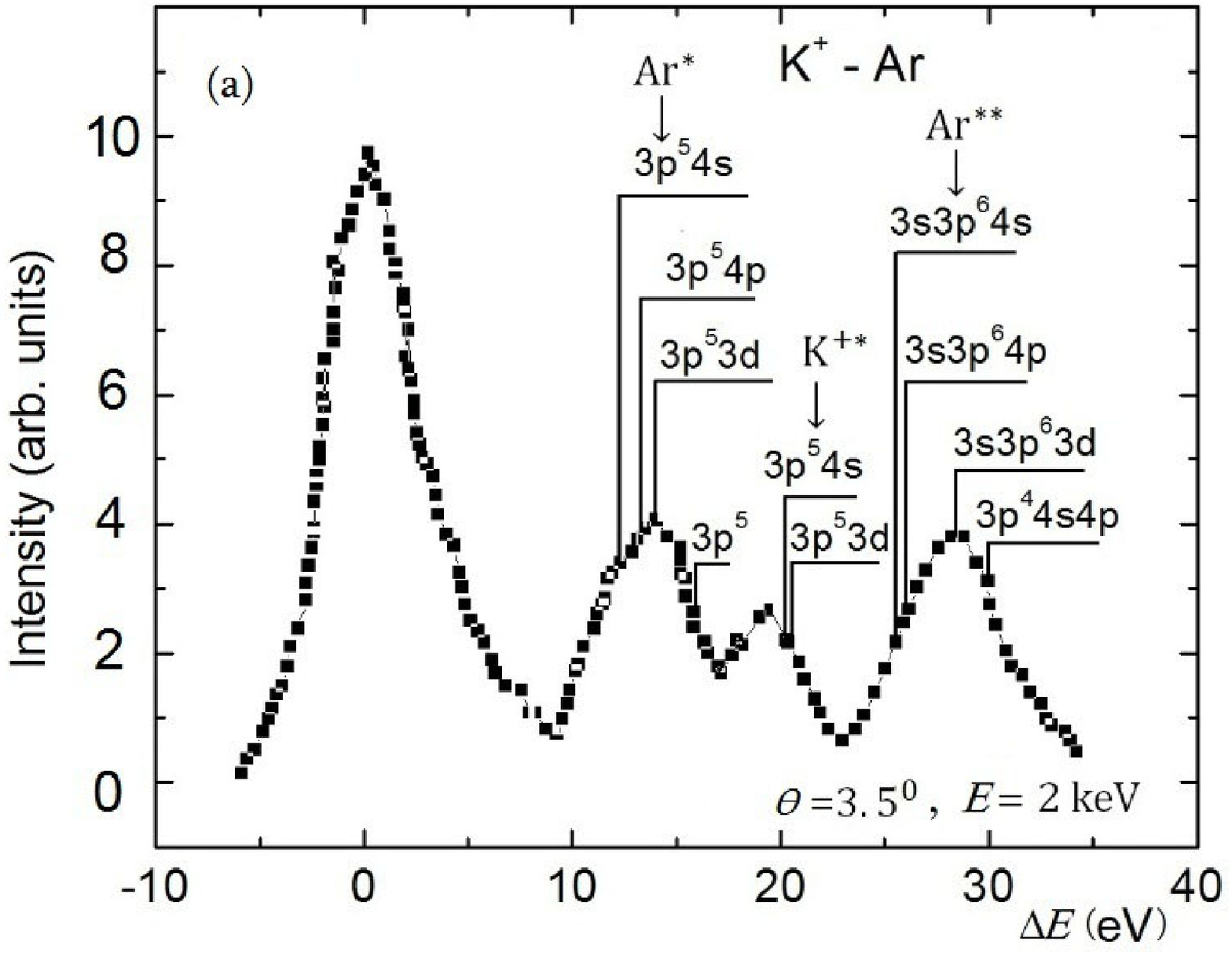} %
\includegraphics[height=6.cm]{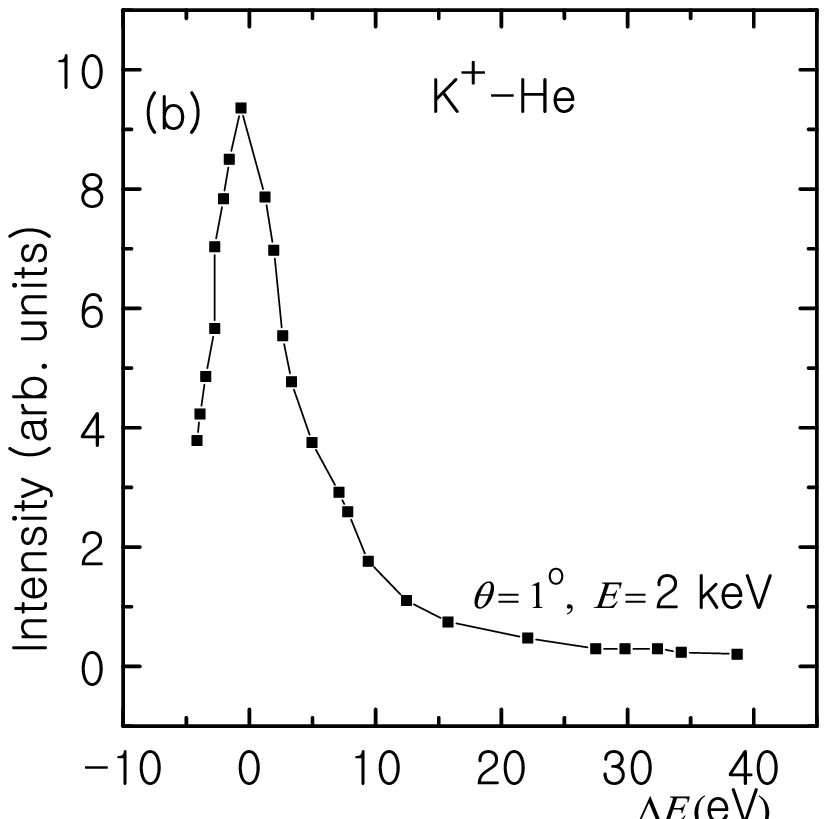}
\par
\hspace{1.cm}
\includegraphics[height=6.cm]{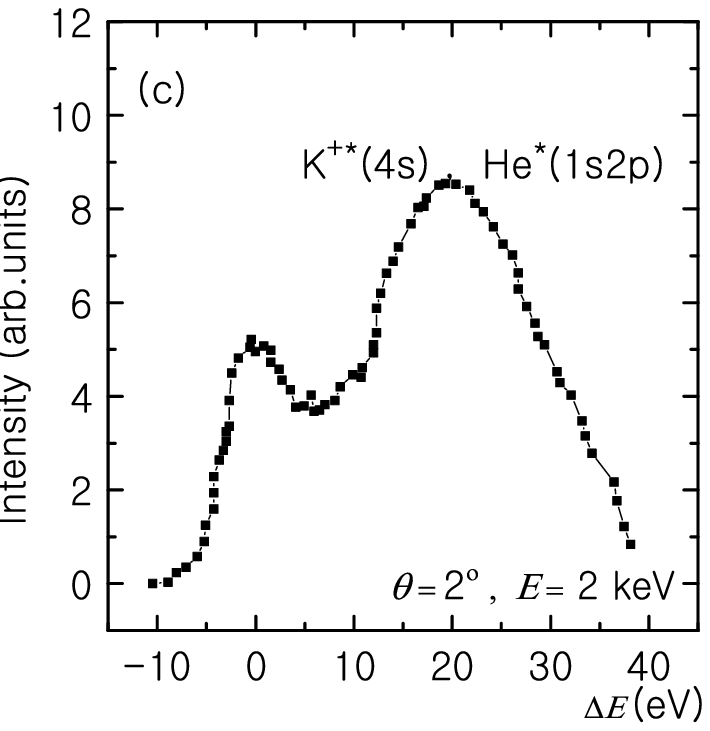} 
\hspace{-1.5cm} %
\includegraphics[height=6.cm]{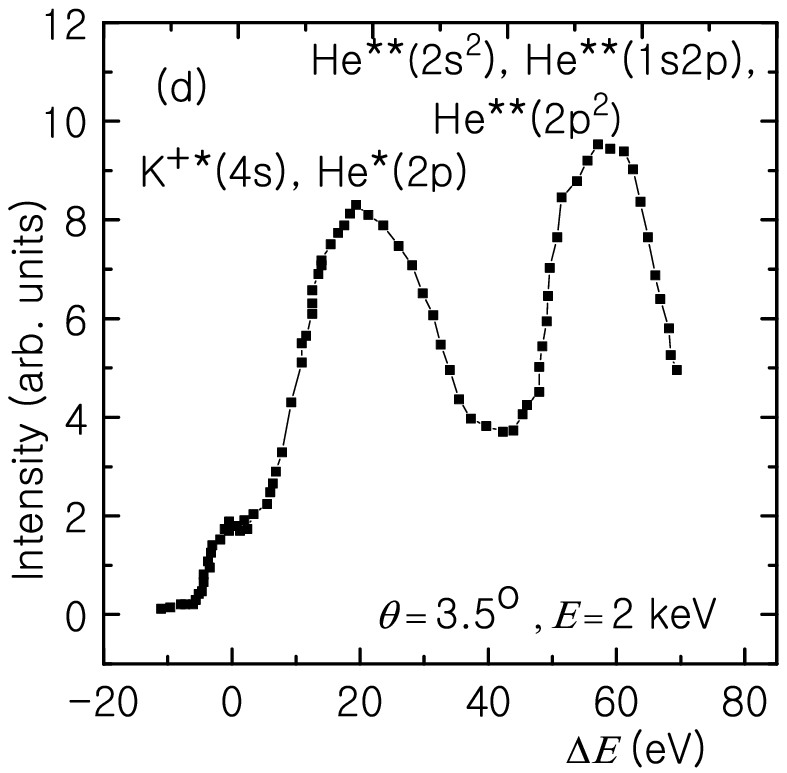}
\end{center}
\caption{Energy loss spectra. (a) The energy loss spectrum for K$^{+}$- Ar
collision. (b)-(d) Energy loss spectra for K$^{+}$- He collisions at
different scattering angles.}
\label{F5}
\end{figure}

Since the mass of projectile K$^{+}$ is much larger than that of the He
target, a significant decrease in the velocity of the relative motion of the
particles when traversing the pseudocrossing region occurs at significantly
large initial ion energies (of the order of 1 keV), unlike for systems with
comparable masses. In this situation, the only particles which can approach
each other to a distance at which the terms undergo a pseudocrossing are
those for which the impact parameters are much smaller than this distance.
The small value of the charge exchange cross section in K$^{+}-$He
collisions compared to those for collisions of certain other pairs of
particles, e.g., K$^{+}-$Ar, K$^{+}-$Kr, K$^{+}-$Xe, at ion energies
0.7-10.0 keV, can be attributed precisely to this effect \cite{Kikiani3}.
For these pairs of particles, the mass of the incident particle is less than
or roughly equal to the mass of the target particle. Therefore, there is no
significant decrease in the relative velocity compared to the initial
velocity of the ion. Correspondingly, there is no significant decrease in
the impact parameters at which the pseudocrossing regions are reached.

\begin{figure}[t]
\centering {\ \includegraphics[width=0.7\textwidth]{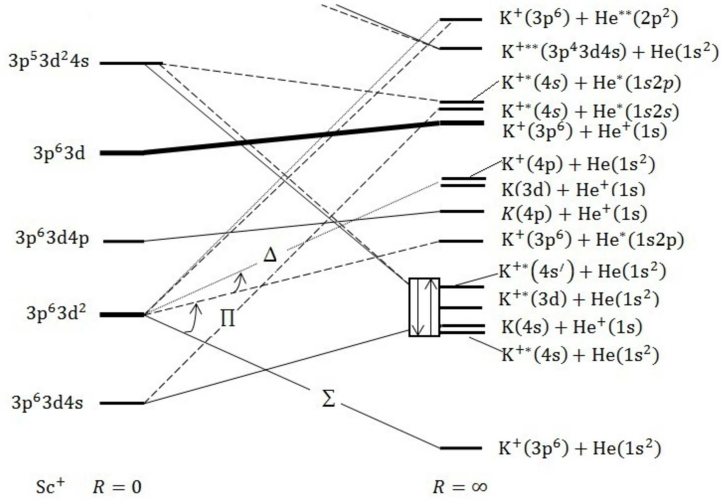} }
\caption{Schematic correlation diagram.}
\label{Fig6}
\end{figure}

In the two-channel approximation, the charge exchange cross section should
increase with increasing of ion energy, approaching the maximum value $%
\sigma _{max}\sim $ 0.5$\pi R_{0}^{2}$, where $R_{0}$ is the position of the
pseudocrossing region \cite{Nikit}. Using a limiting value of 65 eV as the
interaction energy of the particles in the pseudocrossing region, and the
potential curve for the ground state of K$^{+}-$He system from Ref. \cite%
{Nikulin}, we find $R_{^{0}}$ $\approx $ 0.7 \AA\ and correspondingly, $%
\sigma _{max}\approx $ 0.7$\times $10$^{-16}$ cm$^{2}$. It can be seen from
Fig. 2 (curve 3), that the cross section actually increases (up to $\sigma
=5.5\times 10^{-18}$ cm$^{2}$) with increasing of the ion energy only up to
1.5 keV. As the energy further increases, the cross section decreases; and
it increases again only at energies greater than 5 keV. This behavior of the
cross section indicates that at the ion energy of 1.5 keV the charge
exchange mechanism is not correlated exclusively with the interaction of the
entrance and exit channels, since the interference of these channels with
other channels becomes important. In particular, the decrease of the cross
section at ion energies 1.5$-$5.0 keV can be related to the interference of
the exit channel with the channel asymptotically dissociating into to the
state K$^{+}$(3p$^{5}$4s$^{\prime }$)$+$He(1s$^{2}$), as follows from Fig.
6. An interaction of these channels is confirmed by the agreement of the
position of the minimum in the energy dependence of the charge exchange
cross section with the position of the maximum in the energy dependence of
the cross section for the decay of the resonant level of the ion K$^{+}$(3p$%
^{5}$4s$^{\prime }$), shown in Fig. 2. The position of the maximum value of
the cross section 1.5$\times $10$^{-17}$ cm$^{2}$ (see Fig. 2, curve 1)
corresponds roughly to that of the dip in the charge exchange cross section.

\subsection{The ionization process}

It follows from the results of present measurements of emitted electrons in K%
$^{+}-$He collisions and, in part, from the data obtained in Ref. \cite%
{Aizawa}, where the electron spectrum was measured over the interval 5-24
eV, that the liberation of slow electrons (with energies less than 10-15 eV)
is a characteristic of ionization (\ref{ioniz}) in K$^{+}-$He collisions. In
order to determine the channel and mechanism of ionization, we estimate the
contribution of several inelastic processes that result in emission of slow
electrons. The contribution of the direct ionization K$^{+}$(3p$^{6}$)$+$%
He(1s$^{2}$)$\rightarrow $K$^{+}$(3p$^{6}$)$+$He$^{+}$(1s)$+$e, that in the
quasimolecular model is linked to the transition of the adiabatic term into
the continuum, is estimated following Ref. \cite{Solov}. The expression
given in Ref. \cite{Solov} is written in terms of the principal quantum
number for a hydrogen-like atom. Therefore, we modify it slightly for the
estimation of the cross section for the emission of electrons from
multielectron atoms. Then the final expression can be written as:

\begin{equation}
\sigma =2\pi vk\frac{\mid R_{nl}\mid ^{2}}{E_{nl}}{}\text{{Im}}R_{nl}\exp
(-2E_{nl}{}\text{{Im}}R_{nl}/v),  \label{sigma}
\end{equation}%
where $v$ is the relative velocity of the particles in the nonadiabatic
region, $k$ is the number of electrons within the united atom in the state
with quantum numbers ($n,l)$, $E_{nl}$ is the binding energy of an electron
in the nonadiabatic region, and Im$R_{nl}$ and Re$R_{nl}$ are the
coordinates where the potential surfaces cross in the complex plane. They are

\begin{equation}
{}\text{{Im}}R_{nl}=\frac{\sqrt{2l(l+1)}}{Z_{eff}},\text{ \ \ Re}R_{nl}=%
\frac{l(l+1)}{Z_{eff}}\text{ \ \ }(m=0),  \label{RR}
\end{equation}%
where $Z_{eff}$ is the effective charge of the nucleus of the united atom.
The last formula was derived by the same approach as the expression (26) in
Ref. \cite{Solov}, except that the population of the initial orbital is
taken into account, and the binding energy is introduced as a parameter.

\qquad Analysis of the correlations of the molecular orbital in the (KHe)$%
^{+}$ system shows that in the united atom limit, the 1s electrons of the He
atom become 3d electrons of the Sc$^{+}$ ion. Since this result is of
importance for evaluating the contribution of the direct ionization, we note
that it agrees both with the Barat- Lichten correlation rules \cite{Barat}
and the Eichler$-$Wille rules \cite{Eichler}. Consequently, the value $l=2$
was chosen for the evaluation of the cross section. The binding energy $%
E_{nl}$ of the electrons in the nonadiabatic region was chosen to be equal
to that of the 3d electrons of the Sc$^{+}$ ion. The effective charge $%
Z_{eff}$ was determined by the interpolation of the data of Ref. \cite%
{Hartree}. For the 3d electrons of the Sc$^{+}$ ion, the value $Z_{eff}=5.5$
was found. Estimates of the cross section for the direct ionization with
these parameters show that at an ion energy of 2.5 keV, the contribution of
the direct ionization to the total ionization cross section is less than
0.1\%, while at 6.5 keV it is less than 5.5\% i.e., this contribution is
insignificant over the entire energy range studied.

\qquad The double ionization of the He atom, and capture accompanied by
ionization of the He atom evidently make a small contribution to the
ionization cross section. There are two reasons for this: the large energy
defect for these processes, 79 and 74.6 eV, respectively, and the absence of
pseudocrossings of the corresponding quasimolecular terms with the
ground-state term, as follows from the diagram in Fig.6.

As follows from the electron spectrum in Ref. \cite{Aizawa}, the resultant
intensity of the discrete lines associated with capture to the
autoionization states of the K atom and ultimately corresponding to the
ionization of the He atom is several times lower than the that of the lines
of the K$^{+}$ ion corresponding to stripping processes. Since the stripping
cross section is less than 10\% of that for ionization (see Fig. 3) we find
that the capture to autoionization states of K atoms also plays no important
role in the ionization. Consequently, by systematically evaluating the
contributions of various inelastic processes to the ionization of the target
atoms in K$^{+}-$He collisions, we find that the ionization may be caused
primarily by the decay of quasimolecular autoionization states. These could
be expected to be states with two excited electrons (see the energy loss
spectrum in Figs. 5c and 5d), since such states undergo quasimolecular decay
with a high probability, liberating mainly slow electrons with a continuous
energy distribution \cite{Ogur2}. Both of these circumstances agree with the
conclusions which follow from an analysis of the correlation diagram of the
system, and they add a few refinements to the ionization mechanism.

It can be seen from the diagram that the ground state of the system goes
over in the limit of the united atom into the 3p$^{6}$3d$^{2}$ ($^{1}$D$_{2}$%
) singlet state of the Sc$^{+}$ ion. Below this state, the Sc$^{+}$ ion has
only three triplet states (not shown in the diagram) and one singlet state
(with the configuration 3p$^{6}$3d4s) \cite{Radtsig}. Only the term,
asymptotically dissociating into K (3p$^{6}$4s) $+$ He$^{+}$, with the
appropriate $^{1}\Sigma $ symmetry, crosses the ground state. Note that
electron transitions occur primarily between terms of the same multiplicity.
We find that the autoionization results from the filling of primarily terms
of $^{1}\Pi $ and $^{1}\Delta $ symmetry. These terms correlate in the limit
of separated atoms to the states: K$^{+\ast }$(4s)$-$He$^{\ast }$(1s2s), K$%
^{+\ast }$(4s)$-$He$^{\ast }$(1s2p), K$^{+\ast \ast }$ (3p$^{4}$3d4s)$-$He(1s%
$^{2}$) and K$^{+}$(3p$^{6}$)$-$He$^{\ast \ast }$(2p$^{2}$) (Fig. 6). All of
these terms, as expected, correspond to two-electron excitations of the
system. Since the ground-state term has the symmetry $^{1}\Sigma $, the
terms are populated as a result of $\Sigma -\Pi $ and $\Sigma -\Pi -\Delta $
transitions, associated with the rotation of the internuclear axis along the
different nuclear trajectories.

\subsection{The stripping process}

According to our estimates, the governing mechanism for the stripping
process (\ref{striping}) is the transition of adiabatic term into the
continuum. The stripping cross section determined by this mechanism was
calculated using Eqs. (\ref{sigma}) and (\ref{RR}). As it follows from the
diagram in Fig.6 the 3p electrons of the K$^{+}$ ion are correlated with the
3p electrons of the Sc$^{+}$ ion. Consequently, in evaluating the cross
section we selected $l=1$, while we took $Z_{eff}$ to be the same as for the
3d electrons of Sc$^{+}$ ion in evaluating the ionization cross section. The
stripping cross section calculated with these parameters is consistent over
the entire energy region (Fig. 3, curve 1) with the experimental cross
section (Fig. 3, curve 2) to within less than 25\%. Although this agreement
is fortuitous to some extent, it does support the conclusion that this is
the governing mechanism for the stripping: K$^{+}$(3p$^{6}$)$+$He(1s$^{2}$)$%
\rightarrow $K$^{++}$(3p$^{5}$ )$+$He(1s$^{2}$)$+e$.

\subsection{The excitation function of K atom and K$^{+}$ ion}

As shown in Fig. 4, the excitation functions of the K atom and K$^{+}$ ion
for 4p lines are similar, apparently indicating a common mechanism of
population. For the interpretation of this result it is expedient to use the
correlation diagram of molecular states of the (KHe)$^{+}$ system from Fig.
6. We emphasize that the term corresponding to the K(3d) state is
responsible for both -- the population of the potassium atom and the ion
into the 4p state. The terms corresponding to the excitation of K(4p) as
well as K$^{+}$(4p) do not have immediate crossing points with the entrance
term K$^{+}$(3p$^{6}$)$-$He(1s$^{2}$). However, both of these terms may be
populated from that corresponding to K (3d) excitation states by means of a
double rotational transition $\Sigma -\Pi -\Delta $ between the entrance
term and that corresponding to the K(3d)$-$He$^{+}$(1s) state. The
excitation state of the term corresponding to K(4p) may be populated from
the term corresponding to the K(3d), via the cascade transition, from level
3d to 4p. As for the population of the term K$^{+}$(4p), because the term K$%
^{+}$(4p)$-$He(1s$^{2}$) and K(3d)$-$He$^{+}$(1s) at large internuclear
distance are energetically closed (the energy defect is 0.4 eV), the
exchange interaction between these states may cause a population of the term
corresponding to K$^{+}$(4p), and hence the identical energy dependences of
the cross sections. The probability of population of the K(3d) level of the
potassium atom itself is very small, due to the small probability of the $%
\Sigma -\Pi -\Delta $ transition (see Fig. 6). Perhaps this is the reason
for the anomalous small excitation cross section of the potassium atom in
K(4p) state and for the K$^{+}$(4p) state. This can be seen from Fig. 2,
curve 2 and Fig. 4, curves 1 and 2, respectively.

\subsection{The energy loss spectrum}

Another conclusion that can be drawn from Fig. 5 is that the energy loss
spectra for both K$^{+}-$Ar and K$^{+}-$He collisions have a discrete
character. The energy loss spectrum for K$^{+}-$Ar \ in Fig. 5 (a) is
presented for comparison and is chosen for the fixed energy ($E=$ 2 keV) and
scattering angle ( $\theta $ =3.5$^{0}$) at which the inelastic channels are
most populated. The first peak of this spectrum corresponds to the elastic
scattering of K$^{+}$ ions, while the second one corresponds to the single
electron excitation of the Argon atom to (4s), (4p) and (3d) states with
energy losses of $11.6<\Delta E<14$ eV. The third one corresponds to the
excitation of K$^{+}$ ions into the (4s) and (3d) states with energy losses $%
16<\Delta E<22$ eV, and the fourth one corresponds to the single and double
autoionization states of Ar with energy losses $25<\Delta E<32$ eV.

The energy loss spectrum for K$^{+}-$ He colliding pairs is given at $E=2$
keV and for the various scattering angles of K$^{+}$ ions: $\theta =1^{0}$,
Fig. 5 (b); $\theta =2^{0}$, Fig. 5 (c); $\theta =3.5^{0}$, Fig. 5 (d)
respectively. This allows us to observe the dynamics of inelastic channels.
Let us analyze the energy loss spectrum for the K$^{+}-$\text{He pair at the
ion energy of }$E=2$\text{ keV and }$\theta =3.5^{0}$ scattering angle shown
in Fig. 5 (d). The first maximum at zero energy loss is attributed to the
elastic scattering of K$^{+}$ ions. The second maximum at $\Delta E=20-21$
eV is ascribed to the excitation of the potassium ion into the (4s$%
^{^{\prime }}$) states and to the direct excitation of the helium atom into
(2p) state. With low probability of excitation ionization and stripping,
with energy losses $\Delta E=24.6$ eV and $\Delta E=31$ eV, respectively,
are favored. Rather higher excitation probabilities are observed at the
energy loss $58<\Delta E<60$ eV and they are ascribed to the two electron
excitation of the helium atom into (2s$^{2})$ and (2p$^{2}$) states \cite%
{Hicks, kita2}.

The remarkable feature of the spectrum from Figs. 5 (b)$-$5 (d) is the rapid
decrease of the elastic channel intensity with scattering angle. This fact
indicates the increasing importance of inelastic channels. In spite of the
fact that the energy loss spectrum has been measured up to 60 eV, a
significant elastic peak was observed only at 1$^{0}$ scattering angle. With
the increase of scattering angle from 1$^{0}$ to 3.5$^{0},$ the excitation
of the inelastic channel (one and two electron excitation of helium atom and
excitation of potassium ion) becomes predominant. We have also observed
these specific features of the energy loss spectrum for other energies of
the incident ions. This allows us to conclude that for K$^{+}-$He collisions
in this energy range, the main inelastic channel is the excitation of the K$%
^{+}$ ion into the (4s$^{^{\prime }}$) state. This result was also confirmed
from our excitation function given in Fig.2, curve 1. Indeed, as seen from
this figure, the intensities of excited K$^{+}$(4s$%
{\acute{}}%
$) lines are more than one order higher than for the rest of the inelastic
channels considered here.

\section{Conclusion}

\label{Concl}

Using refined experimental methods that include a condenser plate method,
angle- and energy-dependent collection of product ions, energy-loss and
optical spectroscopy under the same "umbrella" and a well checked
calibration procedure of the light recording system, we have measured the
absolute values of the charge exchange, ionization, stripping and excitation
cross sections. The correlation diagram of the (KHe)$^{+}$ system has been
employed to discuss the mechanisms for these processes in the K$^{+}$ ion
energy range $0.7-10$ keV.

Charge exchange in K$^{+}-$He collisions mostly occurs through the channel K$%
^{+}$(3p$^{6}$) $+$ He(1s$^{2}$) $\rightarrow $ K(4s)$+$He$^{+}$(1s)
resulting from the capture of the electron to the ground state of the atom
in regions of pseudocrossings of the potential curves of $^{1}\Sigma $
symmetry.

The primary ionization mechanism in K$^{+}-$He collisions is the filling, as
a result of $\Sigma -\Pi $ and $\Sigma -\Pi -\Delta $ transitions, of
quasimolecular autoionization terms and their decay in the region of the
transition into the continuum (in the stage in which a quasimolecule exists).

Stripping in K$^{+}-$He collisions occurs via a mechanism involving a
transition of adiabatic term into the continuum in the region of a non
adiabatic interaction of molecular orbitals with orbital angular moments
which are identical in the limit of the united atom.

A common mechanism of the populations, as well as the small values of
excitation cross sections for the potassium ion and atom into 4p states are
explained. The usage of the collision spectroscopy method has permitted us
to observe exceptionally highly excited states of helium atom in the energy
loss spectrum.

\acknowledgments

The author are grateful to Prof.\ A. M\"{u}ller for valuable and stimulating
discussions.

\end{document}